\documentclass{moriond}

	\usepackage{textcomp}						
	\usepackage[T1]{fontenc}
	\usepackage[utf8]{inputenc}				
	\usepackage[reqno,tbtags]{amsmath}
	\def\beq{\begin{equation}}
	\def\eeq{\end{equation}}
	\def\bea{\begin{align}}
		\def\eea{\end{align}}

\def\Journal#1#2#3#4{{#1} {\bf #2}, #3 (#4)}

\def\NIMA{{\em Nucl. Instrum. Methods} A}

\def\PRL{\em Phys. Rev. Lett.}
\def\PRD{{\em Phys. Rev.} D}

\def\be{\begin{equation}}
\def\ee{\end{equation}}
\def\bea{\begin{eqnarray}}
\def\eea{\end{eqnarray}}

\begin{document}
\vspace*{4cm}
\title{GRAVITY EXPERIMENTS WITH ULTRACOLD NEUTRONS AND THE \textit{q}\textsc{Bounce} EXPERIMENT}

\author{ T. Jenke, G. Cronenberg, M. Thalhammer, T. Rechberger, P. Geltenbort, H. Abele }

\address{Atominstitut, Technische Universit\"at Wien, Stadionallee 2, 1020 Vienna, Austria}
\address{Institut Laue Langevin, 71 avenue des Martyrs, 38000 Grenoble, France}

\maketitle
\abstracts{This work focuses on the control and understanding of a gravitationally interacting elementary quantum system. It offers a new way of looking at gravitation based on quantum interference: an ultracold neutron, a quantum particle, as an object and as a tool. The ultracold neutron as a tool reflects from a mirror in well-defined quantum states in the gravity potential of the earth allowing to apply the concept of gravity resonance spectroscopy (GRS). GRS relies on frequency measurements, which provide a  spectacular sensitivity.
}

\section{Introduction}

\begin{figure}
	\centering
	\includegraphics[width=0.5\linewidth]{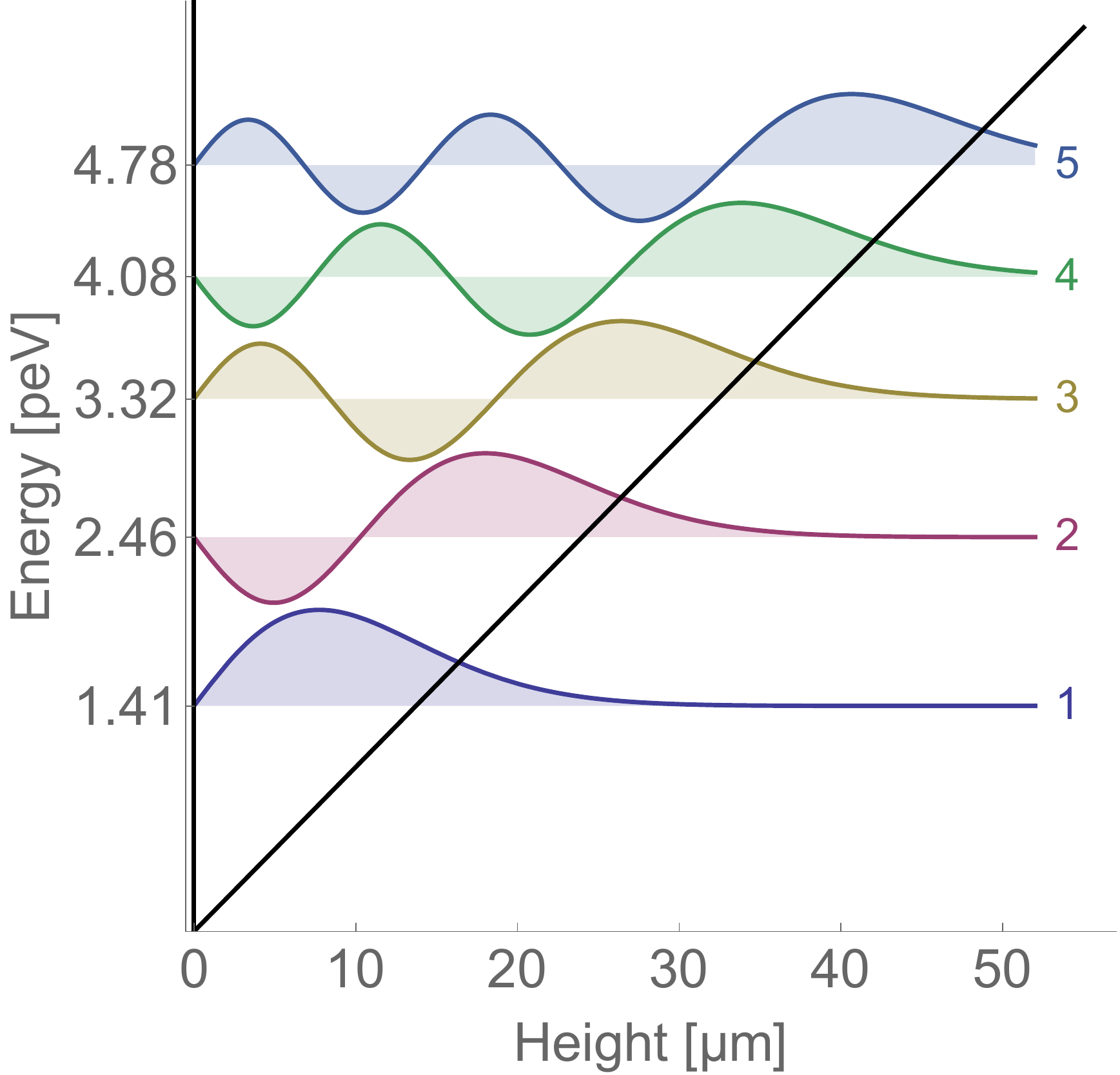}
	\caption{The vertical wave function of the first five eigen states in the gravity potential with their eigen energies. The black lines indicate the potential composed from the neutron mirror and gravity.}
	\label{fig:States-Free}
\end{figure}

We present the quantum bouncing ball: a neutron falling in the gravity potential of the earth and reflecting from a mirror for ultracold neutrons. As typical for bound quantum systems, neutrons have discrete energy eigenstates, and we find them in a coherent superposition of particular levels. The discrete energy levels occur due to the combined confinement of the matter waves by the mirror and the gravitational field. For neutrons the lowest discrete states are in the range of several pico-eVs, opening the way to a new technique for gravity experiments and measurements of fundamental properties. The energy levels together with the neutron density distribution are shown in Fig.~\ref{fig:States-Free}. As Gea-Banacloche\cite{Banacloche1999} has pointed out, the eigenfunctions for this problem are pieces of the same Airy function in the sense that they are shifted in each case in order to be zero at $z = 0$ and cut for $z < 0$, see section~\ref{sec:exp}.

One task is a precise measurement of the energy levels by a resonance spectroscopy technique called Gravity Resonance Spectroscopy (GRS), see section~\ref{sec:grs}. Quantum mechanical transitions with a characteristic energy exchange between an externally driven modulator and the energy levels are observed on resonance. An essential novelty of this kind of spectroscopy is the fact that the quantum mechanical transition is mechanically driven by an oscillating mirror and is not a consequence of a direct coupling of an electromagnetic charge or moment to an electro-magnetic field. The concept is related to Rabi's magnetic resonance technique for the measurements of nuclear magnetic moments.

The other task is  to study the dynamics of such a quantum bouncing ball, i.e. the measurement of the time evolution of such a superposition of quantum states interpreted as reflections, when they come close to the mirror, see section~\ref{sec:qbb}. A quantum mechanical description of UCNs of mass $m$ moving in the gravitational field above a mirror is essentially a one-dimensional (1D) problem. The corresponding gravitational potential is usually given in linear form by $m g z$, where $g$ is the gravitational acceleration and $z$ the distance above the mirror, respectively. The mirror, frequently made of glass, with its surface at $z = 0$ is represented by a constant potential $V_{\mathrm{mirror}}$ for $z<0$. The potential $V_{\mathrm{mirror}}$ is essentially real because of the small absorption cross section of glass and is about 100 neV high, which is large compared to the neutron energy $E$ perpendicular to the surface of the mirror. Therefore it is justified to assume that the mirror is a hard boundary for neutrons at $z = 0$.
Our tasks offer a new way of looking at gravitation based on quantum interference: an ultracold neutron, a quantum particle, as an object and as a tool. This unique system – systematic effects are extremely small – allows to map aspects of gravitation including the dark energy and dark matter searches.

\section{Experiments with Quantum States of Ultracold Neutrons in the gravity field}
\label{sec:exp}
Ultracold neutrons (UCN) bridge the gap between gravity experiments at short distances and the precise measurement techniques of quantum mechanics:
These neutrons are so slow, that their corresponding wave length is much larger than interatomic distances of matter. Hence, they are totally reflected under any angle of incidence (why they are referred to as ultracold), and may bounce on a flat, polished glass mirror. The measurements take place at the Institut Laue-Langevin which houses the worlds most intense steady-mode neutron source. For UCN with a tiny vertical velocity component, quantum mechanics comes into play: Schrödinger's equation with the linear gravity potential tells us, that bound states $\psi_k$ of these UCN with macroscopic size should exist:
\begin{equation}
\left( -\frac{\hbar^2}{2 m_i}\frac{\partial^2}{\partial z^2} + m_g g z \right)  \psi_k
=
i \hbar \frac{\partial}{\partial t} \psi_k.
\label{eq:1}
\end{equation}

Here, $m_i$ and $m_g$ are the inertial and gravitational mass of the neutron, $g$ corresponds to the local acceleration of the earth, and $z$ denotes the height over the glass mirror. The equation can be transformed in order to be dimensionless using a substitution $z \rightarrow z/z_0$, $E \rightarrow E/E_0$, and $t \rightarrow t/t_0$.
The corresponding scaling factors read
\begin{equation}
z_0 = \sqrt[3]{\frac{\hbar^2}{2 m_i m_g}} \approx 5.9~\text{{\textmu}m}, \quad
E_0 = m_g g z_0 \approx 0.6~\text{peV}, \quad
t_0 = \frac{\hbar}{E_0} \approx 1.1~\text{ms}.
\label{eq:2}
\end{equation}
The scaling factors define the typical distance and energy scale of any experiment with gravitationally bound UCN.\\
The solutions of Schrödingers equation are the well-known Airy-functions.
In Fig.~\ref{fig:States-Free}, the first five states are shown. The eigen energies depend solely on the neutron's inertial and gravitational mass, Planck's constant, the zeros of the Airy function and the local acceleration of the earth. Their actual values are in the pico-eV range. Typical sizes of the states are in the range of a few ten microns. Their detection using position-sensitive detectors is feasible.

In order to perform quantum experiments with these states, a state preparation mechanism is needed. One possibility is the introduction of a second neutron mirror with a rough mirror surface, which acts as a boundary condition from top. This new boundary condition has two effects: On the one hand, it influences the shape and eigen energies of the quantum states. On the other hand, it effectively removes higher states from the system, because higher states have a significantly larger overlap with the rough mirror surface, and are in this way scattered off the experiment and absorbed. In fact, quantum states were observed in that way by measuring the surviving neutrons as a function of height of top mirror in a transmission experiment\cite{Nes02,Abele2003,Wes07}. These earlier experiments with neutrons are presented in a review \cite{Abele2008}.

\begin{figure}
	\centering
	\includegraphics[width=0.8\linewidth]{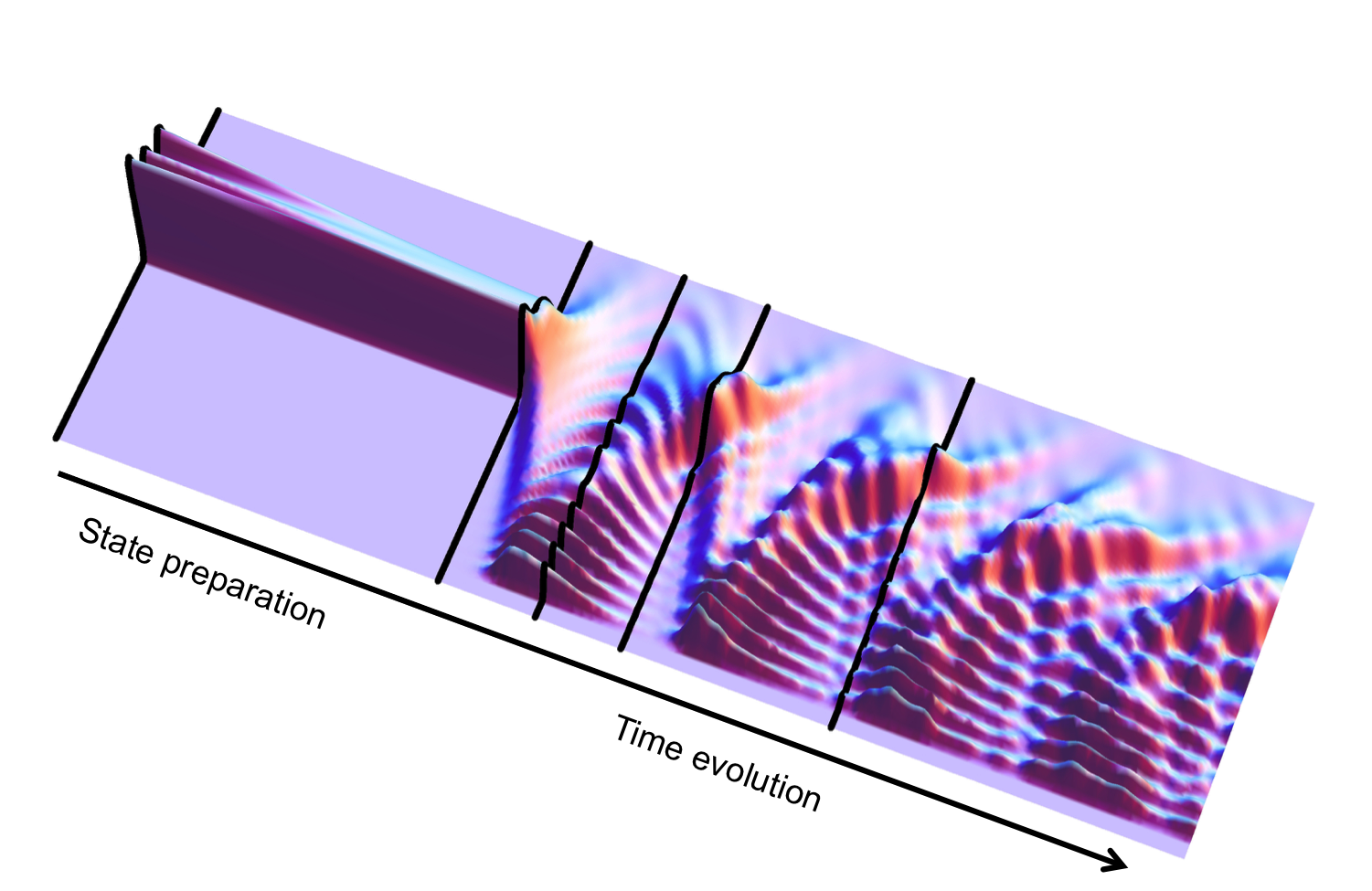}
	\caption{The simulated so-called quantum carpet shows the time-evolution of the wave function of the neutron after falling down the step.}
	\label{fig:Carpet}
\end{figure}

In this article, we present two different kind of experiments with well-prepared wave packets of gravitationally bound quantum states: one might imagine to detect the states itself using a position-sensitive detector with micron-resolution. For this purpose, a well-defined wave packet is dropped a step of a few ten microns. This converts the originally prepared wave packet into a super-position of higher states, which evolve with different timing $e^{-i E_k /\hbar t}$. The time evolution of this so-called Quantum Bouncing Ball (QBB) is shown in Fig.~\ref{fig:Carpet}.
 It can be monitored using position-sensitive detectors. First results of our measurements in 2014 will be presented in section~\ref{sec:qbb}.\\
A second approach takes advantage of the fact that the eigen energies of the states are not equidistant. In fact, the difference in energy of any two states is unique, which allows to treat any two state as effective two-level-system. This principally allows for the implementation of the powerful measuring techniques of quantum physics - resonance spectroscopy. Here, one prepares a wave packet of state~$|p\rangle$. Then, one exposes the system to a periodic perturbation with an oscillation frequency~$\omega$ and oscillation strength~$a$. Using an appropriate oscillation strength, the system will be driven into state~$|q\rangle$ close to the resonance condition~$E_q - E_p \approx \hbar \omega$.
As a third step, the wave packet is analysed with respect to state~$|p\rangle$ using the same mechanism as for the preparation process. The rate of neutrons is recorded versus the oscillation frequency and amplitude. The method converts the measurement of an unknown quantity of energy into a frequency measurement, which can be done with incredibly high precision. In contrast to all other resonance spectroscopy experiments, the described resonance method is not linked to any electro-magnetic force as the transition is induced by mechanical oscillations. However, other groups pursue also electro-magnetically induced transitions\cite{Pignol2014}. We refer to this method as gravity resonance spectroscopy (GRS). The expected sensitivity was outlined in\cite{abele2010}.

\section{Progress on Gravity Resonance Spectroscopy}
\label{sec:grs}
Last Moriond conference, we reported on the first realization of the second experimental technique described in section~\ref{sec:exp}, realized in a simplified setup\cite{Jenke2011,Jenke2011Moriond}:
Here, the state preparation, transition and analysis took place simultaneously. As was shown~\cite{Jenke2011diss}, this simplified setup leads to the same set of Rabi's differential equations, complicated with damping terms. Moreover, the system had to be treated as an effective three-level system, because the transitions $|1\rangle \leftrightarrow |2\rangle$ and $|2\rangle \leftrightarrow |3\rangle$ were close due to the second boundary condition from the rough top mirror. The advantage was a very simple and easy-to-control experimental setup, which is crucial for a successful first-time realization of an experiment.

As these measurements were still statistically limited, they were repeated in two different experiments.
\begin{figure}
	\centering
	\includegraphics[width=0.3\linewidth]{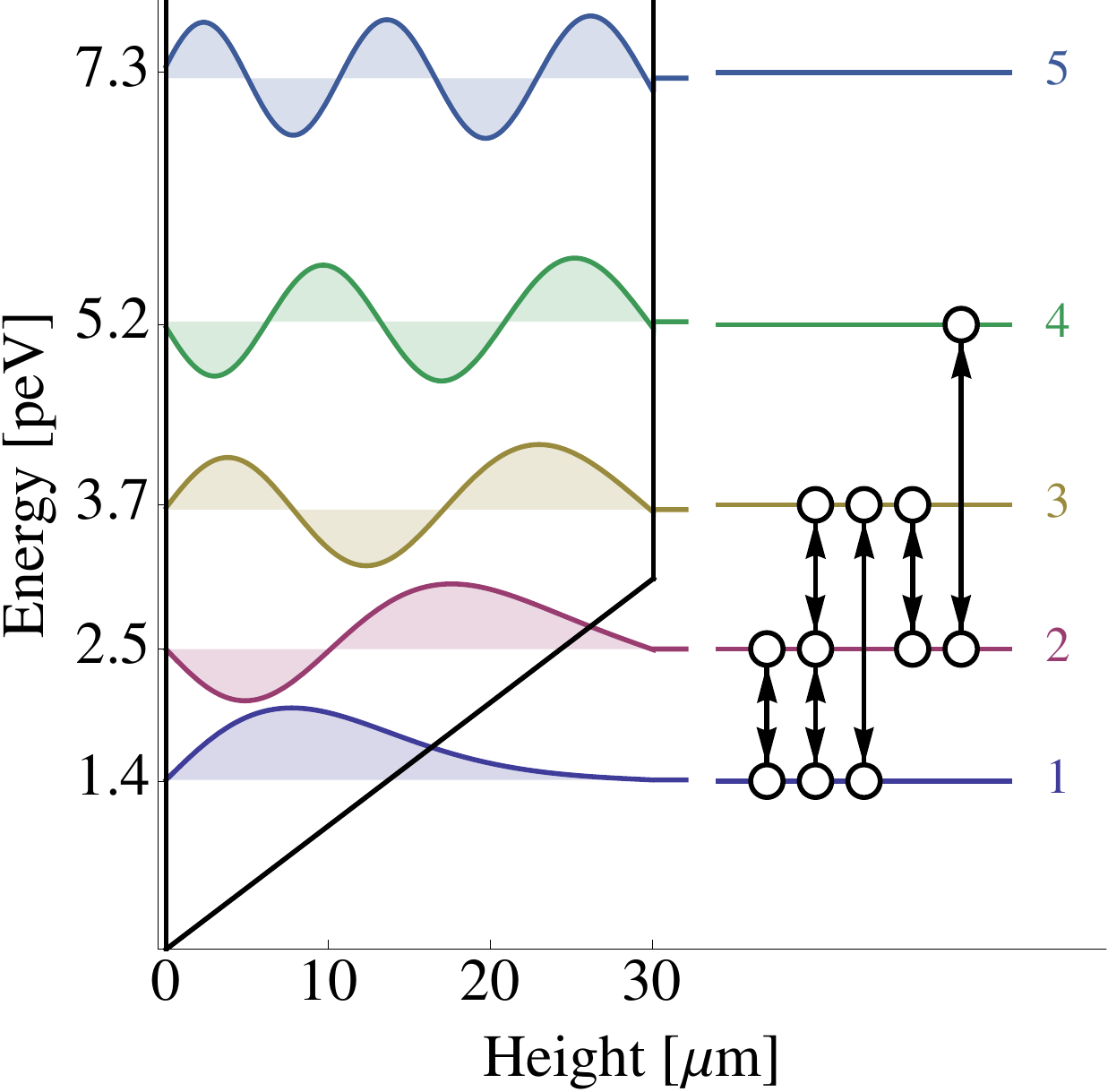}
	\includegraphics[width=0.3\linewidth]{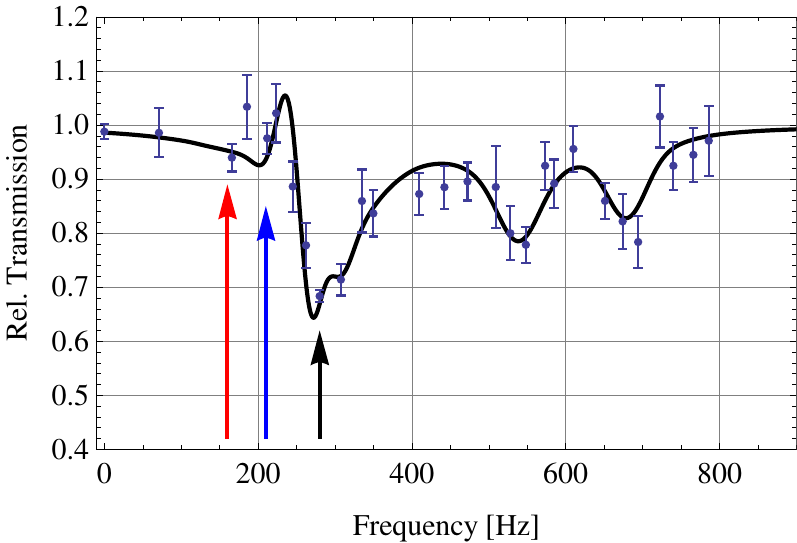}
	\includegraphics[width=0.3\linewidth]{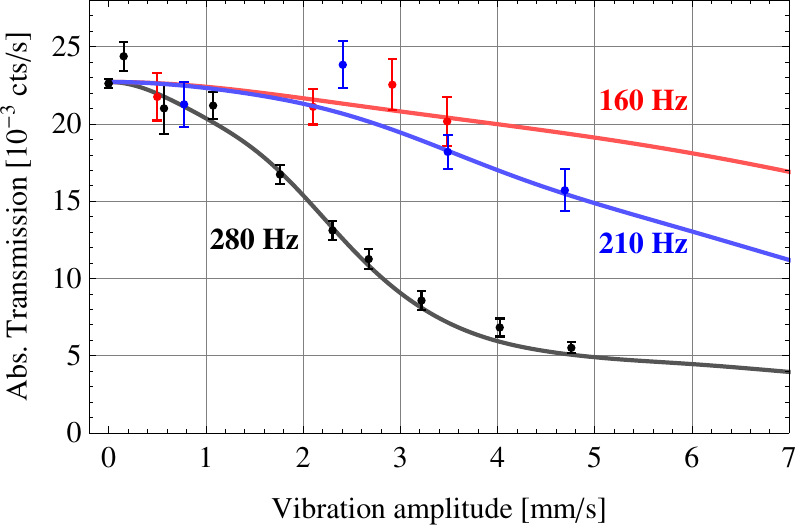}
	\caption{\emph{Left:} The first five wave functions of a neutron confined by a mirror at the bottom and on top separated by 30~{\textmu}m. The many transitions that were driven are indicated by arrows. \emph{Middle:} The transmission at the detector is shown in dependency of the oscillation frequency applied to the system. The rich structure due to the effective  three-level system can be seen. The coloured arrow indicate the frequencies which are shown in the sub-figure to the \emph{right}. \emph{Right:} The transmission decreases with the oscillation amplitude for frequencies close to resonance. Due to the dampening in the system no state revival is observed.}
	\label{fig:transmission-one-part}
\end{figure}
The results are shown in Fig.~\ref{fig:transmission-one-part} and published and described in~\cite{Jen14}. In these measurements, the transitions $|1\rangle \leftrightarrow |3\rangle$, $|2\rangle \leftrightarrow |4\rangle$, and the three-state cascade system $|1\rangle \leftrightarrow |2\rangle \leftrightarrow |3\rangle$ are identified. Moreover, damped Rabi oscillations for the cascade system were studied. The neutron flux dropped close to the resonance condition ($\omega/2\pi \approx 280 Hz$) with respect to the oscillation strength. On the opposite, the flux remained high, when the resonance condition was not fulfilled.

A careful analysis of systematic effects was carried out. The main effect arises due to the rough surface of the upper glass mirror and its influence to the neutrons and its quantum states. While the surface roughness can be measured, its influence on the quantum states is more difficult to predict. The measurements on the roughness were performed using secondary emission microscopy and a mechanical surface roughness scanner, which gave similar results. The influence on the transition frequencies was deduced by large numerical simulations on a super computer.\\
As a result, we concluded that all systematic effects due to the roughness are well below the $10^{-14}$~eV level, which was the level of precision of the experiment. Therefore, we were able to set experimental constraints on the existence of hypothetical Non-Newtonian short-range forces. The experimental limit for the existence of hypothetical chameleon scalar fields improved the existing limit by five orders of magnitude and was the reason for the community of atom interferometry to build dedicated experiments to search for such mechanisms, too. A first result was presented in this Moriond conference\cite{Hamilton2015}.

\begin{figure}
	\centering
	\includegraphics[width=0.6\linewidth]{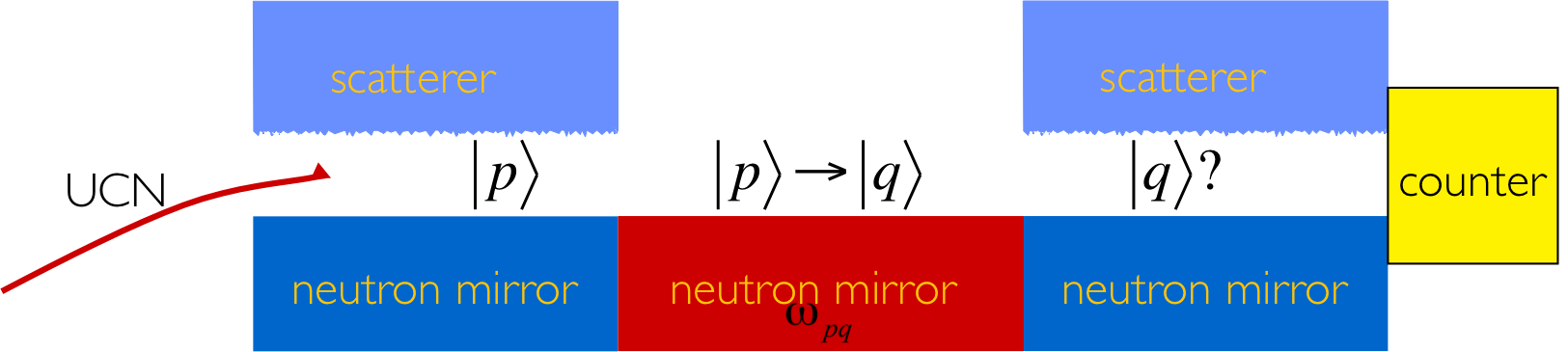}
	\caption{A GRS setup realisation with its three regions for preparation, excitation and analysis. The neutron travels through the setup from left to right and if it survives the setup, it is detected at the neutron counter. Ideally, first the neutron is prepared in the state $\left\vert p\right\rangle$, then the transition to state $\left\vert q\right\rangle$ can be controlled. Finally only the state $\left\vert q\right\rangle$ is which leads to a count-rate drop upon successful transition. In the middle region no upper mirror is present.}
	\label{fig:3-part-setup}
\end{figure}

The next generation GRS experiment we employed, used a-three part realisation of Rabis method (see figure~\ref{fig:3-part-setup}). The three steps preparation, excitation and analysis are now implemented in three physically distinct regions.
The major difference is that in the second region where the excitation between the different quantum levels takes place, no upper mirror confines the neutron from above.
Compared to the one-part setup, no additional energy shift of the states dependant on the slit height occurs, such that the energy of the states only depends on the values mentioned in section two.
This opens up the possibility to determine the inertial and the gravitational mass of the neutron at the same time with the current generation of the experiment\cite{Kajari2010}.

The move to a three part setup came at the cost of increased experimental complexity as more mirrors were used and their alignment to each other needed to be guaranteed.
With improved methods we were able to keep gaps and steps at a level without any influence on the experiment.

\begin{figure}
	\centering
	\includegraphics[width=0.81\linewidth]{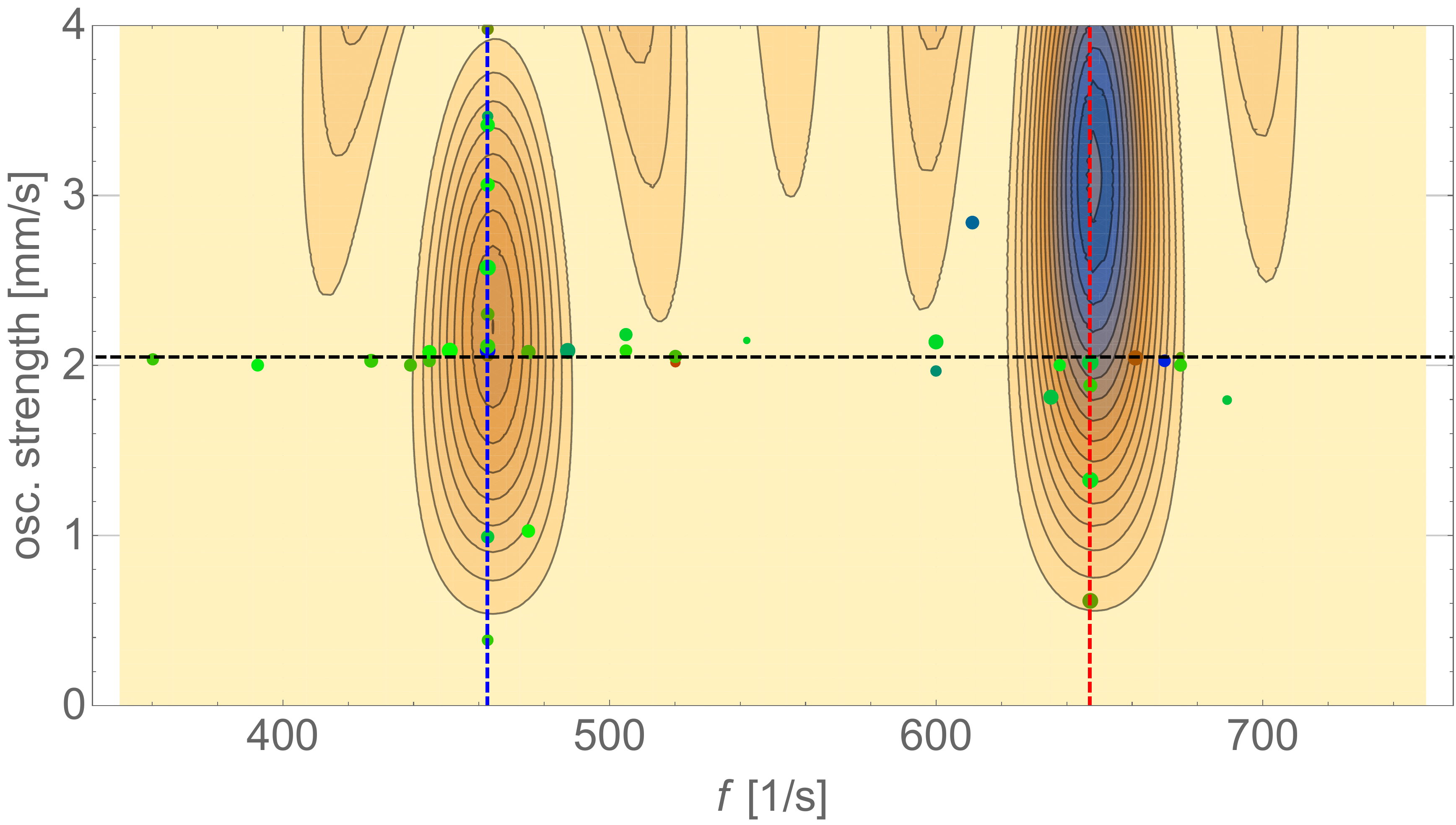}
	\caption{Contour plot of the transmission signal as a function of the oscillation strength and frequency $f$. The coloured dots show the parameters used in the measurements. The colouring of the contour plot show the dips in the transmissions from the transitions $\left\vert 1\right\rangle \leftrightarrow\left\vert 3\right\rangle$ and $\left\vert 1\right\rangle \leftrightarrow\left\vert 4\right\rangle$ at $464.1$~Hz and $648.8$~Hz respectively as obtained by the theory with fitted parameters. The measurements were mainly performed around the same oscillation strength of 2~mm/s. The transitions $\left\vert 1\right\rangle \leftrightarrow\left\vert 3\right\rangle$ and $\left\vert 1\right\rangle \leftrightarrow\left\vert 4\right\rangle$ have been also mapped out for varying oscillation strength.}
	\label{fig:ContourPlot}
\end{figure}

Figure~\ref{fig:ContourPlot} shows all measurement performed in a 2D contour plot.
The theoretical transmission curve with the parameters obtained by a fit for the transitions $\left\vert 1\right\rangle \leftrightarrow\left\vert 3\right\rangle$ and $\left\vert 1\right\rangle \leftrightarrow\left\vert 4\right\rangle$ is plotted as contour.
The dips of the transmission are visible and correspond to successful excitations into the higher state.
The frequencies of the transitions $\left\vert 1\right\rangle \leftrightarrow\left\vert 3\right\rangle$ and $\left\vert 1\right\rangle \leftrightarrow\left\vert 4\right\rangle$ were found at $464.1^{+1.1}_{-1.2}$~Hz and $648.8_{-1.6}^{+1.5}$~Hz respectively. Again, the experiment was statistically limited.

\begin{figure}
	\centering
	\includegraphics[width=0.8\linewidth]{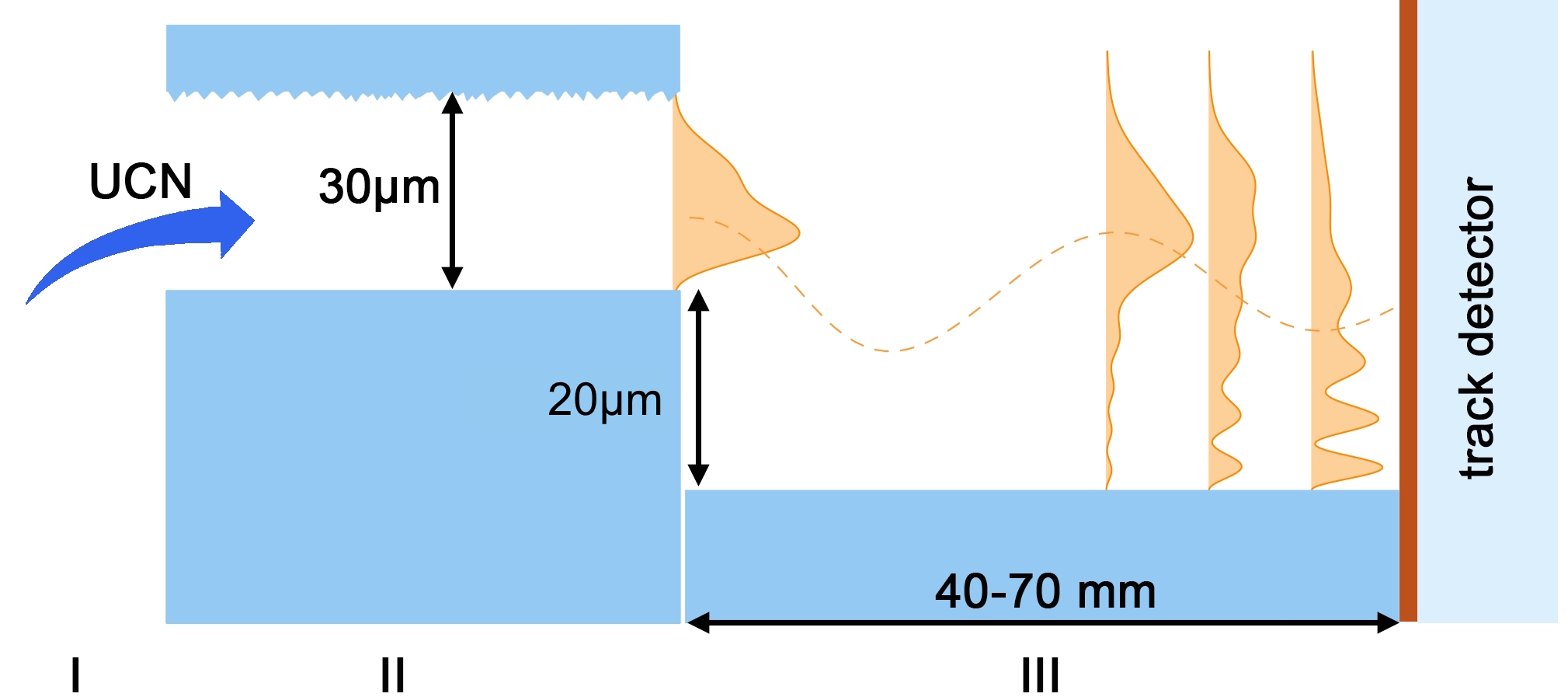}
	\includegraphics[width=0.9\linewidth]{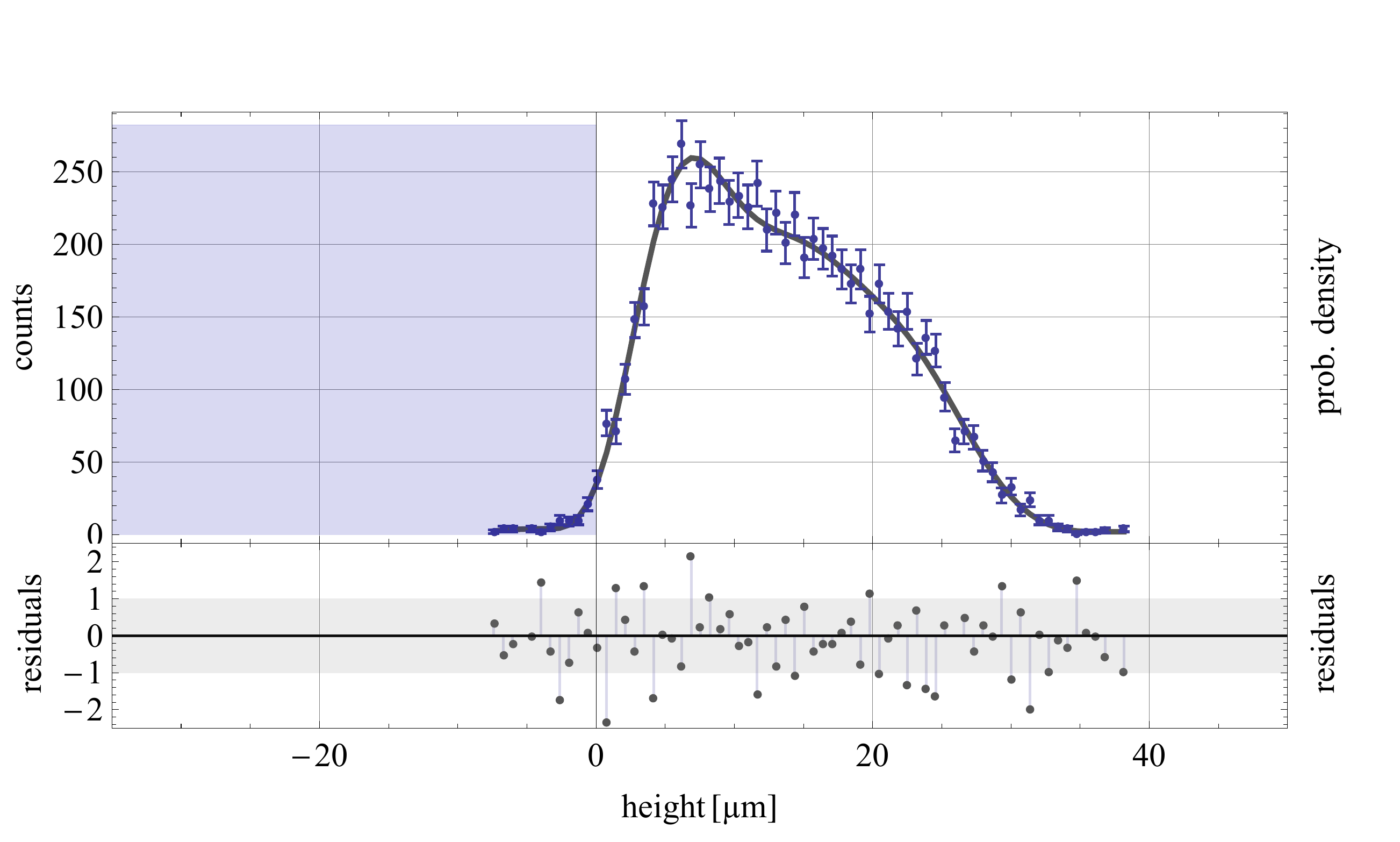}	\includegraphics[width=0.9\linewidth]{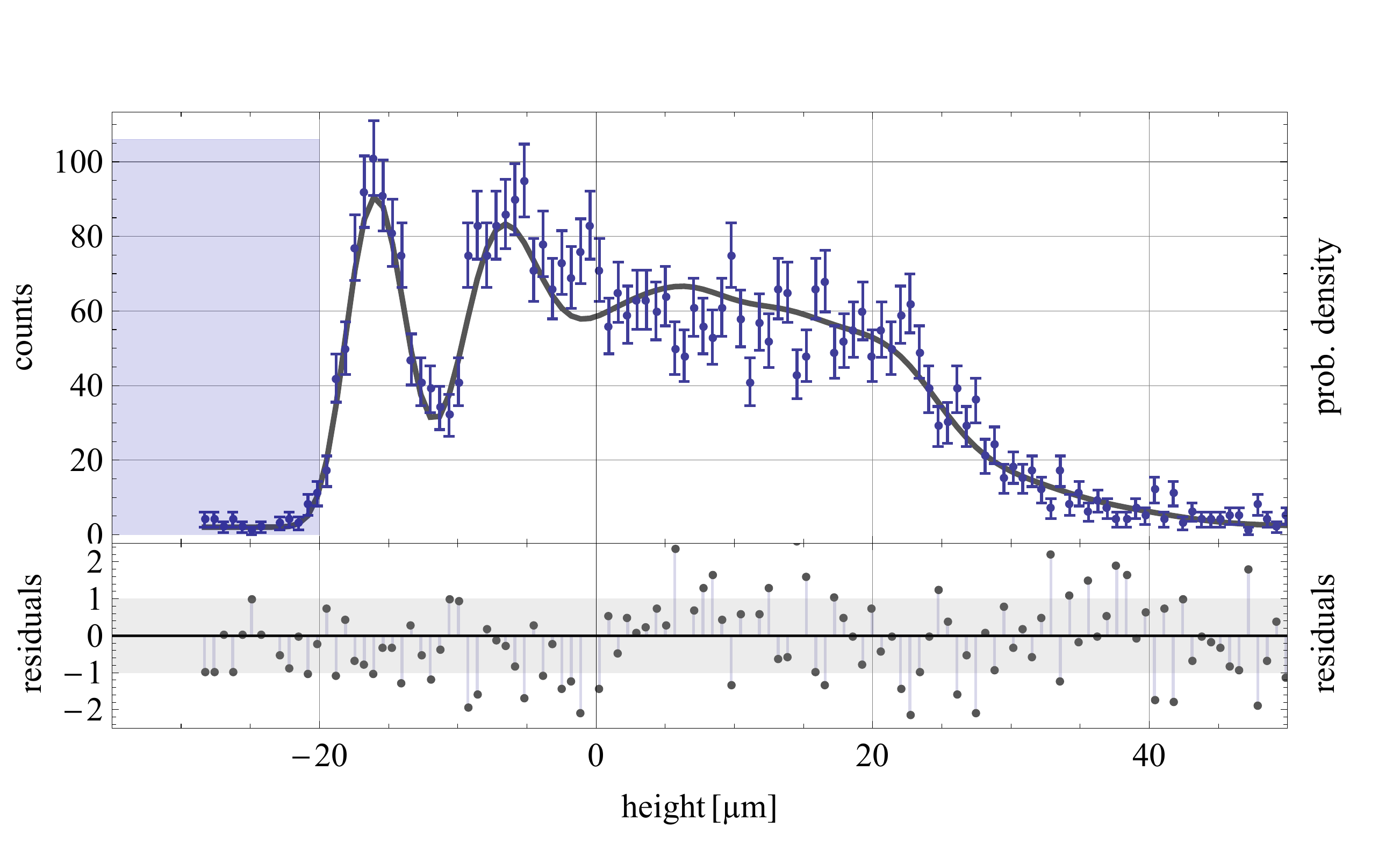}
	\caption{Top: Experimental Scheme. The neutrons pass through the setup from left to right. After a preparation into a superposition of the lowest states, the wave packet is dropped a step of 20~microns and evolves in time. The probability density is recorded using track detectors. Middle/bottom: The spatial probability distribution measured directly at the step and in a distance of 51~mm behind the step. The measurements agree well with the theoretical prediction, a coherent superposition of the gravitationally bound quantum states.}
	\label{fig:QBB-exp}
\end{figure}

From the measured data, improved limits on the hypothetical chameleon field are expected by two orders of magnitude compared to the previous generation.

\section{Realization of a Quantum Bouncing Ball}
\label{sec:qbb}
Gravity resonance spectroscopy offers the possibility to perform very sensitive measurements, because they rely on frequency measurements. These experiments are still strongly limited by statistics, and an improvement by several orders of magnitude seems feasible. Nevertheless, measurements with track detectors are desired, too.
The reason is, that GRS measurements probe the energy scale~$E_0$, while measurements of the spatial probability distribution of the wave packet have access to the distance scale~$z_0$. As can be seen in Eq.~\ref{eq:2}, these scales have a different dependence on the inertial and gravitational mass of the neutron. Therefore, the knowledge of $z_0$ and $E_0$ allows for a test of the weak equivalence principle (WEP) in the quantum regime, which is conceptionally different from ordinary tests. Here, only one quantum particle is used. Moreover, the experiments offer the possibility to study quantum phenomena like quasi-stationary structures in the time evolution of wave packets, so-called quantum carpets, as well as quantum phenomena without classical analoga like collapses and revivals of the wave function. Other so far unobserved aspects of the QBB like collapse and revival of the wave function are presented in \cite{Abele2012}.

These measurements are challenging, because position-sensitive detectors with high efficiency, very low background and a spatial resolution of approx. 1~{\textmu}m are needed. Currently, we use nuclear track detectors with a converter layer of $^{10}$Boron~\cite{Jen13}. First quantum states measurements with time evolution of a coherent superposition are presented in\cite{Abe09,Jen09}. There are several projects~\cite{Jakubek2009,thorsten,japan} under development in order to build "online" detectors, which have the advantage to see the results immediately. A measurement of a neutron spatial density distribution using nuclear track detectors with uranium coating can be found in\cite{Nes05}.

A second challenging point is the implementation of a step between two mirrors, which is precisely known and stable on a level of much less than a micron for a few days, the typical time to take one snapshot. We solve the problem of stability by mounting the mirrors on two nanopositioning tables, that are working in closed-loop operation. The step is measured by capacitive sensors, which directly monitor the mirror surfaces (which are coated with aluminium for that purpose). The sensors are moved over the surfaces using another micropositioning table and monitor the step in this way. The positioning errors of this micropositioning table (roll-, pitch- and  yaw angle) would directly affect the measurement of the step. Therefore, the movement of the table is measured using three additional capacitive sensors that measure the distance to a large measurement plane, which consists of another glass mirror coated with aluminium. In this way, the step was controlled on a stability level of 10~nm.

In 2014, the actual measurements were again carried out at the beam position PF2 at the ILL. Nine snapshots of the Quantum Bouncing Ball were taken within 75 days of beam time.
Here, we present the first two snapshots, see Fig.~\ref{fig:QBB-exp}. An experimental scheme of the experiment can be found in the top figure. Neutrons traverse the setup from left to right. They are prepared into a wave packet of the lowest states in a 30~{\textmu}m wide slit of a flat glass mirror on bottom and a rough one on top. The step was adjusted to either 20~or 30~{\textmu}m. In order to realize different evolution times, the length of the second mirror was adapted.
The figure in the middle shows the height profile of the neutrons directly after the preparation process. It verifies, that the neutrons form wave packets containing only the lower states. The lower figure shows one snapshot of the Quantum Bouncing Ball, taken with a step size of 20~{\textmu}m and a mirror length of the second mirror of 51~mm. This is the position where the expectation value in height should reach its minimum, and the quantum wiggles of the quantum carpet should be at maximal visibility. Indeed, the quantum carpet reaches a contrast of approx.~50\%.

Clearly, the microscope calibration enters the result of the extracted value for the distance scale~$z_0$ directly. Currently, the microscope is calibrated to an accuracy of~1\%. The statistical error on the common data set is well below this value.
Therefore, the data evaluation process and microscope calibration is currently on-going.

\section{Outlook}

It is quite natural to follow the path of history and extend GRS currently using Rabi's method with Ramsey's method of separate oscillating fields\cite{abele2010}. An additional region of state evolution between two oscillating regions will steepen the transmission signal and thus enhance the energy sensitivity of the setup.
This can be used to search for a plethora of new effects. For example a hypothetical electric charge of the neutron might be detectable by the energy shift the different states obtain in an external electric field\cite{Dur11}.
The QBB with improved spatial resolution and systematically rigorous experiments with sufficient count-rate statistic will demonstrate fascinating and simultaneously simple quantum effects.
Combining the results from the QBB and the GRS measurements will allow to test the Weak Equivalence Principle as GRS probes the intrinsic energy scale $E_0$ of the system while QBB measures the length scale $z_0$.

The continuous improvement of the qBounce experiments, both the Quantum Bouncing Ball and Gravity Resonance Spectroscopy show promising potential to tackle the questions about new Fifth Forces.

\section*{Acknowledgments}
We thank D.~Seiler (TU M{\"u}nchen) for preparing the spatial resolution detectors and R.~Stadler (Univ.~of Heidelberg) for preparing the rough mirror surface. We thank T.~Brenner (ILL) for technical support.
We gratefully acknowledge support from the Austrian Fonds zur F{\"o}rderung der Wissenschaftlichen Forschung (FWF) under Contract No.~I529-N20, and No.~I531-N20, No.~P26781-N20, and the German Research Foundation (DFG) as part of the Priority Programme (SPP) 1491. We also gratefully acknowledge support from the French Agence Nationale de la Recherche (ANR) under Contract No.~ANR-2011-ISO4-007-02 and FWF No.~I862-N20, Programme Blanc International -- SIMI4-Physique.

\end{document}